\documentstyle[12pt]{article}

\def\p{\phi}

\def\pr{\partial }
\def\bpr{\bar {\pr }}
\def\ca{{\cal A}}
\def\zb {{\bar z}}
\def\na{{\nabla }}
\def\G{{\Gamma }}
\def\l{{\lambda }}
\newcommand{\be}{\begin{equation}} \newcommand{\ee}{\end{equation}}
\newcommand{\bea}{\begin{eqnarray}}\newcommand{\eea}{\end{eqnarray}}

\begin{document}
\baselineskip= 24 truept
\begin{titlepage}

\title { Hidden Isometry in a Chiral Gauged WZW Model }

\author{Supriya Kar $^{a}$, Alok Kumar $^{a}$ and Gautam Sengupta $^{b}$ \\
\\
a. Institute of Physics, Bhubaneswar-751 005, India. \\
 email: supriya, kumar@iopb.ernet.in\\
\\ b. Center for Theoretical Studies, \\Indian Institute of Science,
	Bangalore-560 012, India.}


\date{}
\maketitle

\thispagestyle{empty}

\vskip .6in
\begin{abstract}

\vskip .2in

\noindent  		It is shown that the asymmetric chiral gauging of the
WZW models give rise to consistent string backgrounds.
The target space structure of the
${[{SL(2,\Re)/ {SO(1,1)}}]}_L \bigotimes {[{SL(2,\Re)/U(1)}]}_R$
model is analyzed and the presence of a hidden isometry in this
background is demonstrated. A nonlinear coordinate transformation
is obtained which transforms the asymmetric model to
the symmetric one, analyzed recently by two of the present authors.

\end{abstract}

\bigskip

\rightline {IP/BBSR/92-67}
\rightline {IISc-CTS/8-92}
\rightline {September,1992}
\vfil

\end{titlepage}

\eject

Various techniques have been recently devised for studying
and generating large number of classical solutions of $[1,2,3]$ string theory.
The solutions obtained by gauging $[4,5,6]$ of the WZW
models have been quite  popular, since they provide an exact
string theory generalization of many of the nontrivial solutions of general
relativity. An exact conformal field theory
describing a two dimensional black hole $[7]$ can be obtained
by gauging the axial or vector $U(1)$ subgroup of an $SL(2,R)$
$WZW$ model $[4]$. Several interesting generalizations
of this model have been reported $[5,6]$.

In general,  the resulting background
obtained by gauging still has residual isometries.
Most often these isometries, like translations,
can be obtained by a simple observation of the background configuration.
It has been shown that
for $d$-isometries, acting as translations on the coordinates,
the string effective
action ($SEA$) is invariant under an $O(d, d)$ group $[1,2]$
of transformations.
These transformations are useful
in relating  those classical solutions with each other
which depend on the same number of coordinates.
These symmetries manifest themselves in terms of the conserved
currents on the string worldsheet.

In this paper, we show that another gauging, namely
the chiral gauging $[8,9,10]$, can be used in left-right asymmetric
manner for the construction of consistent string backgrounds.
We explicitly work out the case of the asymmetric chiral gauged
${[{SL(2,\Re)/{SO(1,1)}}]}_L \bigotimes {[{SL(2,\Re)/U(1)}]}_R$
model. The resulting target space has an isometry acting as translation.
However, unlike the case of the symmetric chiral gauged,
${[{SL(2,\Re)/U(1)}]}_L \bigotimes {[{SL(2,\Re)/U(1)}]}_R$,
model $[9]$, the other translation isometry is absent
and the metric cannot be diagonalized by any coordinate independent
orthogonal transformation. Since the background in the two
cases depend on different number of coordinates, they cannot be
transformed into each other by any global transformation like $O(d, d)$
symmetry mentioned above.

We also show  the presence of a hidden isometry in the background
configuration of the asymmetric chiral gauged
model mentioned above by explicitly finding out the Killing vectors.
The corresponding coordinate
transformations are nonlinear and they give rise to a chiral
conserved current on the worldsheet. Then, by using the tensor
transformation properties of the Killing vectors, we obtain a set of general
coordinate transformations and analytic continuations which transform
the background fields of the asymmetric model to the symmetric one $[9]$.
Interestingly,
the transformation is much more nontrivial than the case of the vector
gauging, where the
choice of different gauged subgroups lead to backgrounds which are
simply related by an analytic continuation.

We start by writing down
the action for a general chiral gauged WZW ($CGWZW$) model $[8]$,
\be
S = S^{WZW} + {k\over 2\pi }\int d^2z\; Tr\; [\; A^{R}_{z}\;
U^{-1}\bar{\pr }U + A^{L}_{\bar z}\; {\pr }U\;
U^{-1} + A^{R}_{z}U^{-1}A^{L}_{\bar z}U\; ] ,
\ee
\noindent where $A^R_z\; (A^L_{\bar z})$ in eq.(1) are the $z (\bar
z)$ components of the gauge fields $A^R_{\mu }\; (A^L_{\mu })$.
$U (z, {\bar z}) \in \ G$ and the gauge fields $A^R_z$ ($A^L_{\zb }$)
correspond to the generators in subgroups $H_1$ ($H_2$) of the group $G$.
The action (1) has an underlying gauge invariance $[8]$ whose
transformation parameters are independent in the left and
right-moving sectors. The action (1)
gives a Lagrangian representation
of the coset conformal field theory ${[{G/ {H_1}}]}_L \bigotimes
{[{G/ {H_2}}]}_R$ where $H_1$ and $H_2$ are two subgroups of $G$
which are not necessarily same and may even have different ranks.

We work out the case when the group $G$ in eq.(1) is $SL(2,R)$, $H_1
= SO(1, 1)$ and  $H_2 = U(1)$.
For our example, therefore, the full nontriviality of the
chiral gauging is not incorporated, since both $H_1$ and $H_2$ have
the same rank. This however allows us to illustrate another
interesting aspect, namely that they are related to the symmetric
chiral gauging, i.e., $H_1 = H_2 = U(1)$ by a nonlinear coordinate
transformation and analytic continuation.

For the asymmetric chiral gauged model, the currents corresponding to
the gauged subgroups are
$J^L_{\bar z} \equiv -{i\over 2}\; Tr({\sigma }_1 U^{-1}\pr _{\bar z}
U)$ in the left and
$J^R_{z} \equiv\; {1\over 2}\; Tr(\sigma _2\pr _zU\; U^{-1})$ in the
right. We represent the gauge fields as $A^L_{\bar z} \equiv
\big ({1\over 2}{\ca }^L_{\bar z}\sigma _1\big )$, $A^R_z \equiv
\big (-{i\over 2}{\ca }^R_z\;\sigma _2\big )$ and
parametrize the group manifold as
$U\equiv exp\; {\big ({i\over 2}\p _L\;\sigma _2\big )}\; exp\;
{\big ({r\over 2}\;\sigma _1\big )}\; exp\; {\big ({i\over 2}\p _R\;
\sigma _2\big )}$. Then, by integrating out the gauge fields $[4]$
from the action (1), we get
\bea
&&S={k\over 8\pi } \int d^2z\;\big[ \pr r\bar\pr r
 - {\pr }{\p }_L{\bar\pr }{\p }_L
+ \pr {\p }_R{\bar\pr }{\p }_R
+2\coth r\cot {\p }_L \pr r\bar\pr {\p }_L\nonumber\\
&&\qquad\qquad\qquad
+2{\cot {\p }_L\over {\sinh r}} \pr r\bar\pr {\p }_R
+R^{(2)}(\ln \sin {\p }_L \sinh r + const.)\big] .
\eea

\noindent The background fields are therefore
found to consist of the metric ${(G_{\mu\nu })}$,
antisymmetric tensor ${(B_{\mu\nu })}$ and dilaton ($\Phi$):
\be
(G + B) =\left (\matrix {{1} & {2\coth r\cot {\p }_L} &
{{2\cot {\p }_L}/ {\sinh r}}\cr {0} & {-1} & {0}
\cr {0} & {0} & {1}\cr }\right )
\ee
\noindent and $\Phi = - \ln \sin {\p }_L \sinh r + const.$
Here we have modded out a constant $(k/4)$.

We observe that the above background fields depend on two coordinates
($r$ and ${\p }_L$).
We have explicitly verified that the above background configuration
satisfy the field equations of the three dimensional
$SEA$ for the cosmological constant $V=-1$.
The scalar curvature is
$R = -[{7/ {(2\sin^2 {\p }_L}\sinh^2 r )}]$.
This shows that the target space-time contains a
line singularity along $\p_L\ = 0$ and $\ \pi $ for all $r$.
However unlike the case of the three dimensional
black string $[9]$, the translation symmetry along the singularity direction is
lost. In the asymptotic limit,
the metric becomes flat and the torsion
vanishes. We have also checked that the underlying $O(1,1)$
symmetry of the $SEA$ for the background (3), cannot generate any
inequivalent solution.

We now examine the symmetries of the background configuration (2).
In this connection, it
has been shown that there is a one-to-one correspondence between the
chiral conserved currents on the worldsheet
and the set of Killing vectors (labeled by an index $i$)
satisfying the equations $[11]$,
\be
{K_i}^{\mu } {\na }_{\mu }{\Phi }\; =\; 0
\ee
\be
{\pr }_{\mu }{K_i}_{\nu }\ -\ {{\G }^{\l }}_{\mu\nu }
K_{i\l }\ \pm\ {1\over2} {H^{\l }}_{\mu\nu }
K_{i\l }\ =\ 0
\ee
\noindent where the plus (minus) signs before the torsion term in eq.(5)
correspond to the left (right)-chiral currents $i.e.
\ \pr {\bar J} = 0 $ ($\bpr J = 0$). The isometry direction is given by the
coordinate transformation, $\delta {x^{\mu }} = \epsilon {K_{i}^{\mu }}$.

The translation in $\p _R$
is obviously a symmetry of the background in eq.(3)
and gives rise to the right-chiral current,
$J = \Big (\pr {\p }_R + {{\cot {\p }_L}\over {\sinh r}}\pr r\Big )$.
This  corresponds to the Killing vector,
$K_1^{\mu} \equiv (K_1^r,\ K_1^{\p_L},\ K_1^{\p_R}) = - (0, 0, 1)$.

We now show the presence of a hidden isometry for the background in eq.(3).
This isometry acts nonlinearly on the space-time coordinates
and gives a new chiral conserved current on the worldsheet. We will
show that, in fact these are the  only isometries of the background
in eqs.(2)-(3) which can give chiral conserved currents.

To find the new hidden isometry, we start by writing
the most general candidate for the left-chiral current as,
\be
\bar J\ =\ f{\bpr r}\ +\ g{\bpr }{\p }_L\ +\ h{\bpr }{\p }_R
\ee
\noindent where $f$, $g$ and $h$ are apriori arbitrary functions of $r$, $\p_L$
and $\p_R$. We then impose the condition that
\bea
&&\pr \bar J = 0 = -(\sinh r\sin^2 \p_L )\ (\ [E_{\p_R}]\ g_{,\ r} + [E_{\p_L}]
                   \ h_{,\ r} + [E_r]\ h_{,\ \p_L})
                         \nonumber\\
	            &&  + f_{,\ \p_L}\pr\p _L \bpr r + f_{,\ \p _R} \pr \p_R
				   \bpr r + g_{,\ \p _R}\pr\p_R \bpr \p_L
					+ h_{,\ \p_R} \pr\p_R\bpr\p_R \nonumber\\
			    &&  + [A]\ \pr r\bpr r + [B]\ \pr \bpr r +
					[C]\ \pr \p_L \bpr\p_L + [D]\ \pr\bpr\p_L
					+ [F]\ \pr\bpr\p_R \ ,
\eea

\noindent where $E_{\mu } = {\delta S}/{\delta {x^{\mu }}}$,
$(x^{\mu } = r,\ \p_L,\ \p_R)$ are the coordinate variations of the
action in eq.(2) and vanish due to the field  equations.
Explicit form for the quantitities, [A], [B], [C], [D] and [F] are given as,
\bea
&&[A] = f_{,\ r} - g_{,\ r}\sin \p_L\cos \p_L\coth r + h_{,\ r}
        \; ({{\sin \p_L\cos \p_L}/ {\sinh r}}) \nonumber\\
&&[B] = f + g_{,\ r}\sin \p_L \cos \p_L - h_{,\ r}\sin \p_L\cos \p_L\cosh r
        + h_{,\ \p_L}\sin^2 \p_L\sinh r \nonumber\\
&&[C] = g_{,\ \p_L} - h_{,\ \p_L}\cosh r \nonumber\\
&&[D] = g + h_{,\ r}\sin^2 \p_L\sinh r + h_{,\ \p_L}\cos\p_L\sin\p_L\cosh r
                       \nonumber\\
&&[F] = h + g_{,\ r}\sin^2\p_L\sinh r + h_{,\ \p_L}\sin\p_L\cos\p_L \ .
\eea
\noindent Using the equations of motion, $E_{\mu } = 0$,
we now conclude that $\pr {\bar J} = 0$ gives the conditions
$ f_{,\ \p_L} = f_{,\ \p_R} = g_{,\ \p_R} = h_{,\ \p_R} = 0 $
and	$[A] = [B] = [C] = [D] = [F] = 0$.
This implies a set of five first order partial differential
equations. After some manipulations, it can be shown that they have a
unique solution given by
$ f(r) = 0$ , $\; g(r,\p_L) = ({{C_0}\coth r}/ {\sin {\p_L}})\; $ and
$\; h(r,\p_L)\ =\ ({C_0}/ {{\sin {\p }_L \sinh r}})\; $,
where $C_0$ is a constant. Therefore the expression for the
only left-chiral current is given as,
\be
\bar J\ =\ \big ({C_0}/ {{\sin \p_L \sinh r}}\big )
\big (\cosh r\ {\bpr}\p_L\ +\ {\bpr }\p_R\big ) .
\ee
\noindent The corresponding Killing vector is written as,
\be
	K_2^{\mu} \equiv (K_2^r, K_2^{\p_L}, K_2^{\p_R})
	= i\ [\cos {\p }_L ,\ (-\sin {\p }_L \coth r),
	   \ ({\sin {\p }_L / {\sinh r}})]
\ee
and it can be checked that $K_{2}^{\mu }$ satisfies the Killing eqs.(4) and (5)
for the background in eqs.(2)-(3).

By similar technique, one can also show that our model has $J$
as the only right-chiral conserved current.
The presence of the extra Killing vector $K_2^{\mu}$ implies a larger
symmetry in the target  space.
We believe that such hidden isometries may be present in much wider
class of classical string solutions.  It will be interesting to
investigate the presence of these isometries in other backgrounds so
as to understand their possible moduli deformations. For example, in
this paper we now show the connection of our solution
to another one, namely the symmetric chiral gauged model $[9]$, which has
the same number of left and right-chiral currents.

The symmetric chiral gauged
${[{SL(2,\Re)/ U(1)}]}_L \bigotimes {[{SL(2,\Re)/U(1)}]}_R$
model is analyzed in ref.[9]. It is represented  by
the nonlinear sigma model action,
\be
S^S={k\over 8\pi } \int d^2z\ [{\pr }{\theta }
{\bar\pr }{\theta } + \pr\p {\bpr\p} + \pr \rho\bpr\rho
+ {2\over {\cosh \rho}}\pr\theta \bpr\p
+ {R^S}^{(2)}\ (ln\ \cosh \rho + Const.)]
\ee
\noindent where $\rho $, $\theta $ and $\p $ are the target space coordinates.
This action describes the three dimensional static charged
black string with scalar curvature $R^S = ({7/ {2\cosh^2 \rho }})$
when the same factor $(k/4)$ as in eq.(3) is once again modded out.
The only chiral conserved currents for the symmetric case are,
${\bar J}^s = (\bpr\theta + {1\over {\cosh \rho }}\bpr\p )$
and $J^s = (\pr\p + {1\over {\cosh \rho }}\pr\theta )$.
They correspond to the isometries acting as translation in
$\theta $ and $\p $ respectively. The Killing vectors associated with those
isometries are simply ${\tilde K}_{1}^{\mu } = (0,1,0)$ and
${\tilde K}_{2}^{\mu } = (0,0,1)$.

In order to examine whether the backgrounds in eqs.(2) and (11) are related
by a coordinate transformation, we find it useful to use the
tensor transformation  property of a Killing vector
$K_{i\mu } =({{\pr {\tilde x}^{\nu }}/ {\pr x^{\mu }}})
{{\tilde K}_{i\nu } ^i}\ ,\; (i = 1,2) $, where
$K_{i\mu }$, ${\tilde K }_{i\mu }$ are the
covariant components of the Killing vectors.
Then by writing these equations explicitly in the components of
$K_{i\mu }$, we obtain a set of linear algebraic equations in
($\theta _{, r},\ \p_{, r}$), ($\theta _{,\p_L},\ \p_{,\p_L}$)
and ($\theta _{,\p_R},\ \p_{,\p_R}$).
We also use the invariance of the scalar curvature which
implies, $\cosh \rho = a_0\ \sinh r\sin \p_L$, for some constant $a_0$.
The above set of algebraic equations then have a unique solution,
which can be integrated to give,
\bea
&&\rho\ =\ \cosh^{-1}\big (i\ \sin {\p }_L \sinh r\big )\nonumber\\
&&\theta\ =\ i\ \tanh^{-1}\big (\cos {\p }_L \tanh r \big )\nonumber\\
&&\p \ =\ \tan^{-1}\big (\tan {\p }_L \cosh r \big )\ +\ {\p }_R \ .
\eea

We have explicitly
verified that the background metric in the two cases, namely the
asymmetric  and the symmetric models are related by
the coordinate transformations (12). For comparison, we would like to
mention that for the vector or axial gauging$[4]$, choice of gauging
$SO(1,1)$ or $U(1)$ is related simply by an analytic
continuation of the time coordinate. Further, unlike our case, the
vector or axial models do not allow independent left-right gauging.

We expect that the presence of the new isometry may give rise to a
coordinate dependent  generalization of
the global $O(d,d)$ symmetry of the string effective action,
since it acts more nontrivially than just the translation.
Similar techniques may be employed to
search for higher spin currents on the worldsheet and they may have
interesting target space interpretation.

\vfil\eject

\noindent {\bf Acknowledgements:}\hfil\break
We are grateful to A.P. Balachandran for a very interesting suggestion.
We would also like to thank A. Ali, C. Aulakh, S.P. Khastgir,
J. Maharana and S. Mukherji for discussions.

\def\np{Nucl.Phys.\ {\bf B}}
\def\pl{Phys. Lett.\ {\bf B}}
\def\pr{Phy.Rev.}
\def\prl{Phys. Rev. Lett.}
\def\ml{Mod.Phys.Lett.\ {\bf A}}

\baselineskip 12pt
\vfil
\eject

\end{document}